\documentclass [12pt]{article}
\begin{document}

\title{\textbf{Commutation Technique\\for
interacting close-to-boson excitons}}
\author{O. Betbeder-Matibet and M.
Combescot \\ \small{\textit{GPS, Universit\'e
Denis Diderot and Universit\'e Pierre et Marie
Curie, CNRS,}}\\ \small{\textit{Tour 23, 2
place Jussieu, 75251 Paris Cedex 05, France}}}
\date{}
\maketitle

\begin{abstract}

The correct treatment of the close-to-boson
character of excitons is known to be
a major problem. In a previous work, we have
proposed a ``commutation technique'' to
include this close-to-boson character in their
interactions. We here extend this technique
to excitons with spin degrees of freedom as
they are of crucial importance for many
physical effects. Although the exciton total
angular momentum may appear rather appealing
at first, we show that the electron and hole
angular momenta are much more appropriate
when dealing with scattering processes. As an
application of this commutation technique to
a specific problem, we reconsider a previous
calculation of the exciton-exciton scattering
rate and show that the proposed quantity is
intrinsically incorrect for fundamental
reasons linked to the fermionic nature of the
excitons.
\end{abstract}

\vspace{2cm}


\newpage

Non-linear effects in the optical properties
of semiconductors have received considerable
interest, both experimentally and
theoretically. In these non-linearities, the
interactions between carriers play a crucial
r\^{o}le. Up to now, two types of theoretical
methods have been proposed to deal with these
interactions. 

One method starts with the exact
semiconductor hamiltonian written in terms of
the free electron and free hole fermion
operators $a_{\mathbf{k}}$ and
$b_{\mathbf{k}}$, and ends with the so-called
semiconductor Bloch equations $^{(1)}$, or
better, with an elaborate hierarchy of
2n-point density matrices $^{(2-5)}$ which are
then dynamically truncated to a given order
in the radiation field. These approaches
basically lead to a set of coupled equations
for the time evolution of some expectation
values of these \emph{free electron and free
hole} fermion operators. As the Bloch
equations originally rely on the Hartree Fock
approximation, various extensions have been
proposed to include some correlation effects.
In
particular, it is possible to recover the low
excitation regime, in which the free
electron-hole pairs give rise to exciton
resonances, by dropping all non-linear terms
denoted as exchange processes $^{(1)}$.
However, being written in terms of
free electron-hole pairs, these procedures are
obviously appropriate to systems well
represented by \emph{free} pairs. Even if
various correlation effects can be
included, these methods are mainly
suitable at high density, when the screening
is such that the \emph{exact} correlations
making the excitons are not crucial. For
dilute electron-hole plasma however, these
approaches can appear as starting from the
``wrong'' side, in the sense that, the
excitons being the proper states at low
density, they not immediately appear as the
zero order terms. It is a
priori rather subtle to extend these
procedures \emph{with complete security} to
the low density regime in which the excitons
play the dominant r\^{o}le.

The other method seems much more appealing at
first in the low density regime, since it
relies on the fact that, in this limit, the
electrons and the holes are bound into
excitons \nolinebreak; as these excitons are
made of two fermions, they very much look like
bosons due to the spin statistics theorem.
This idea leads to the so-called bosonic
method
$^{(6)}$, in which the exact excitons are
replaced by boson-excitons and the exact
hamiltonian by an effective hamiltonian
$H_{\mathrm{eff}}$ in which appear
boson-exciton operators. In addition to an
obvious free exciton energy part, this
effective hamiltonian must contain the
interactions between excitons which are
dominant at low density. Besides a ``direct''
term, which corresponds to Coulomb
interactions between excitons made with the
same couples of electrons and holes, $(e,h)$
and $(e',h')$, it has been shown $^{(7)}$
that the interaction part of
$H_{\mathrm{eff}}$ also contains an
``exchange'' Coulomb term which corresponds to
Coulomb interactions between excitons made
with different couples of electrons and
holes, $(e,h)$ and $(e',h')$ on one side,
while $(e,h')$ and $(e',h)$ on the other
side. It is said that this exchange term has
to be introduced in order to take into
account the fermionic character of the
excitons, i.\ e.\ the fact that the excitons
are made of two fermions which can be coupled
in different ways. We have been amazed to
realize $^{(8)}$ that this exchange term,
quoted by everyone for 25 years, is incorrect
since it induces a non-hermitian part in the
effective exciton hamiltonian. Even if this
dramatic error is fixed, which is always
possible at least intuitively, the major
trouble with this bosonic approach actually
comes from the fact that it misses purely
fermionic terms $^{(9)}$ : \linebreak Indeed
the fermionic character of the exciton does
not appear through Coulomb exchange terms
only.

This fermionic aspect is indeed very subtle,
and it is necessary to use a ``full-proof''
procedure to derive all its consequences
properly. The previous approaches, which all
\nolinebreak
$^{(10)}$ end with incorrect results, were too
intuitive to pick up all the consequences of
this fermionic aspect properly.

In short, we can say that :

(i) On one hand, the
semiconductor Bloch equations and their
extention through 2n-point density
matrices use ``clean'' free electron and
free hole fermion operators. They have however
problems with Coulomb interaction and its
resulting correlations, which suffer from
truncation procedures (even if they are
somehow ``controlled''). Indeed this
Coulomb interaction must be included exactly
in order to properly handle the evolution of
the exciton bound states, i.\ e.\ the poles in
the response function, which dominate at
small density. 

(ii) On the other hand, the bosonic
method introduces the exact exciton states a
priori, but uses ``too clean'' boson
operators to represent these excitons, so
that it misses some important consequences of
their underlying fermionic structure.

What we would really like to do is to work
with exact excitons a priori, i.\ e.\
electron-hole pairs in which the Coulomb
interaction responsible for the bound states
is put \emph{exactly}, while we properly
handle the fact that these excitons are not
real bosons. This is this approach we are now
proposing. It has already been briefly
reported in ref.\ (8), without the spin
degrees of freedom for simplicity.

The paper is organized as follows :

In part I, we recall some very basic results
on excitons in semiconductors to settle the
notations.

In part II, we introduce the Coulomb
creation operator
$V_i^\dag$ \emph{between} the exciton $i$ and
the rest of the system and show how we can
get out of it a \emph{direct} Coulomb
scattering $\xi_{lnij}^{\mathrm{dir}}$.

In part III, we introduce the boson-departure
operator $D_{ij}$ and the bare exchange
coefficient $\lambda_{lnij}$ associated to it.

In part IV, we show how we can produce
various Coulomb exchange scatterings out of
$\xi_{lnij}^{\mathrm{dir}}$ and
$\lambda_{lnij}$.

In part V, we use our commutation technique
to calculate the matrix elements of the
\emph{exact} semiconductor hamiltonian
between two-\emph{exact}-exciton states.

In part VI, we reconsider a previous approach
$^{(11)}$ to the exciton-exciton scattering
rate in which enter these $H$ matrix elements
and show why it cannot be correct.

\section{Exact semiconductor hamiltonian and
one-exciton states}
The semiconductor conduction electrons
correspond to S states with a $s=\pm1/2$ spin
along an arbitrary $z$ direction. Let us call
$a_{\mathbf{k},s}^\dag$ the creation operator
for such a free conduction electron with
momentum $\mathbf{k}$ and spin $s$. For the
free valence electrons, the situation is much
more complex : According to the
Kohn-Luttinger representation $^{(12)}$, the
valence energy matrix $^{(13)}$ is given by
\begin{equation}
H_{\mathbf{k}}=Ak^2+B(\mathbf{k}.
\mathbf{I}_{3/2})^2\ ,
\end{equation}
if one neglects warping, $\mathbf{I}_{3/2}$
being the $3/2$ angular momentum matrix. The
$H_{\mathbf{k}}$ diagonalization, which is
done by taking the $z$ axis along the
$\mathbf{k}$ direction, generates the
so-called light and heavy valence electrons.
If we now add Coulomb interaction, we have
shown \nolinebreak $^{(14)}$ that this Coulomb
interaction is not diagonal between heavy and
light valence electrons, a fact which
seriously complicates the problem since it
mixes the
$m=\pm3/2$ with the $m=\pm1/2$ valence states.
In quantum wells, one can forget about this
complexity because the $\pm3/2$ and $\pm1/2$
valence states having different hole masses,
they are shifted differently by the
confinement, so that the heavy-light hole
Coulomb coupling usually gives a negligible
contribution due to the splitting energy
induced by the confinement, which appears in
denominators. In bulk material however, the
only consistent way to forget about these
heavy-light hole Coulomb couplings is to
assume $B=0$, which corresponds to take all
the valence electrons with the same mass. In
doing so, $H_{\mathbf{k}}$ is diagonal for
any $\mathbf{k}$ direction, so that the
valence electron states are simply
characterized by a quantum number $m=\pm3/2,
\pm1/2$, along a direction which is now
arbitrary. Let us call
$b_{\mathbf{k},m}^\dag$ the creation operator
for such a free hole with momentum
$\mathbf{k}$ and angular momentum $m$.

The exact electron-hole semiconductor
hamiltonian then reads
\begin{equation}
H=H_0+V_{\mathrm{eh}}+V_{\mathrm{ee}}
+V_{\mathrm{hh}}\ ,
\end{equation}
where
\begin{equation}
H_0=\sum_{\mathbf{k},s}(\Delta+
\epsilon_{\mathbf{k}}^e)
a_{\mathbf{k},s}^\dag\,a_{\mathbf{k},s}+
\sum_{\mathbf{k},m}\epsilon_{\mathbf{k}}
^h\,b_{\mathbf{k},m}^\dag\,
b_{\mathbf{k},m}\ ,\hspace{1cm}\epsilon_
{\mathbf{k}}^e=\frac{\hbar^2k^2}
{2m_e}\hspace{1cm}\epsilon_
{\mathbf{k}}^h=\frac{\hbar^2k^2}
{2m_h}\ ,
\end{equation}
$\Delta$ being the band gap. For quantum
wells, $m=\pm3/2$, while for bulk materials
\linebreak
$m=\pm3/2,\pm1/2$, as a unique hole mass
$m_h$ is then assumed for the two
hole bands. Within this approximation, the
Coulomb scattering of a hole is diagonal with
respect to its ``spin'' $m$, so that a hole
$(\mathbf{k},m)$ is scattered into
$(\mathbf{k}+\mathbf{q},m)$ with the
\emph{same} $m$. The hole-hole, electron-hole
and electron-electron interactions thus read
\begin{eqnarray}
V_{\mathrm{hh}}&=&\frac{1}{2}\sum_
{\mathbf{q}\neq\mathbf{0},\mathbf{k},\mathbf{k'},m,m'}
V_{\mathbf{q}}\,b_{\mathbf{k}+\mathbf{q},m}^\dag
\,b_{\mathbf{k'}-\mathbf{q},m'}^\dag
\,b_{\mathbf{k'},m'}\,b_{\mathbf{k},m}\ ,
\\V_{\mathrm{eh}}&=&-\sum_
{\mathbf{q}\neq\mathbf{0},\mathbf{k},\mathbf{k'}
,s,m}
V_{\mathbf{q}}\,a_{\mathbf{k}+\mathbf{q},s}^\dag
\,b_{\mathbf{k'}-\mathbf{q},m}^\dag
\,b_{\mathbf{k'},m}\,a_{\mathbf{k},s}\ ,
\\V_{\mathrm{ee}}&=&\frac{1}{2}\sum_
{\mathbf{q}\neq\mathbf{0},\mathbf{k},\mathbf{k'},s,s'}
V_{\mathbf{q}}\,a_{\mathbf{k}+\mathbf{q},s}^\dag
\,a_{\mathbf{k'}-\mathbf{q},s'}^\dag
\,a_{\mathbf{k'},s'}\,a_{\mathbf{k},s}\ ,
\end{eqnarray}
where $V_\mathbf{q}=\frac{4\pi
e^2}{\mathcal{V}q^2}$ in 3D and
$V_\mathbf{q}=\frac{2\pi
e^2}{\mathcal{S}q}$ in 2D, with $\mathcal{V}$
being the sample volume and $\mathcal{S}$ the
well area.

From the free electron and free hole
operators $a_{\mathbf{k},s}^\dag$ and
$b_{\mathbf{k},m}^\dag$, one can construct
the exciton operators $B_i^\dag$ as
\begin{equation}
B_i^\dag\equiv B_{\nu_i,\mathbf{Q}_i,s_i,m_i}
^\dag
=\sum_{\mathbf{k}_i}\langle\mathbf{k}_i|
x_{\nu_i}\rangle\,a_{\mathbf{K}_i^e,s_i}^\dag\,
b_{\mathbf{K}_i^h,m_i}^\dag\ ,
\end{equation}
where we have set
\begin{equation}
\mathbf{K}_i^e=\mathbf{k}_i+\alpha_e\mathbf{Q}_i
\ ,\hspace{1.5 cm}\mathbf{K}_i^h=-\mathbf{k}_i
+\alpha_h\mathbf{Q}_i\ ,\hspace{1.5 cm}
\alpha_{e,h}=\frac{m_{e,h}}
{m_e+m_h}\ .
\end{equation}
$\mathbf{Q}_i$ and $\nu_i$ are the $i$
exciton center of mass momentum and relative
motion quantum parameter : In 3D, $\nu_i$
corresponds to the quantum numbers
$(n_i,l_i,m_i)$, while in 2D it corresponds
to two quantum numbers only $(n_i,m_i)$.
$\langle\mathbf{k}|x_{\nu_i}\rangle$ is the
$i$ exciton relative motion wave function in
$\mathbf{k}$ space. In a similar way, it is
possible to write the free electron-hole pair
operator in terms of exciton operators :
\begin{equation}
a_{\mathbf{k}_e,s_i}^\dag\,b_{\mathbf{k}_h,m_i}
^\dag
=\sum_{\nu_i}\langle
x_{\nu_i}|\alpha_h\mathbf{k}_e-
\alpha_e\mathbf{k}_h\rangle\,
B_{\nu_i,\mathbf{k}_e+\mathbf{k}_h,s_i,m_i}
^\dag\ ,
\end{equation}
as checked by inserting Eq (9) into Eq (7).
Let us stress that this sum contains the
bound states as well as the diffusive states.

The one-exciton state $B_i^\dag|v\rangle$,
where $|v\rangle$ is the electron-hole vacuum
state, is eigenstate of $H$. Using Eqs (2-9),
and the following Schr\"{o}dinger equation for
the electron-hole relative motion in
$\mathbf{k}$ space,
\begin{equation}
(\epsilon_\mathbf{k}^e+\epsilon_\mathbf{k}^h)
\langle\mathbf{k}|x_{\nu_i}\rangle-\sum_{
\mathbf{q}\neq\mathbf{0}}V_\mathbf{q}\,\langle
\mathbf{k}+\mathbf{q}|x_{\nu_i}\rangle=
\epsilon_{\nu_i}\langle\mathbf{k}|x_{\nu_i}
\rangle\ ,
\end{equation}
where $\epsilon_{\nu_i}$ is the energy of the
relative motion state $|x_{\nu_i}\rangle$, it
is indeed easy to check that
\begin{equation}
H\,B_i^\dag|v\rangle=(H_0+V_{\mathrm{eh}})B_i
^\dag|v\rangle=E_i\ B_i^\dag|v\rangle,
\hspace{1.2 cm}E_i=\Delta+\epsilon_{\nu_i}+
\frac{\hbar^2\mathbf{Q}_i^2}{2(m_e+m_h)}\ .
\end{equation}
Let us again stress that in 3D this result
strongly relies on the fact that the Coulomb
interaction is diagonal between holes, i.\
e.\ it does not contain terms in
$b_{\mathbf{k'}-\mathbf{q},m'}^\dag
b_{\mathbf{k'},m}$ with $m'\neq m$.

In many problems dealing with excitons, the
photons play an important r\^{o}le. As a
$\sigma_+$ photon with spin $(J=1,M=\pm1)$
creates an exciton with the same \emph{total}
angular momentum, it may appear appropriate
to use, instead of the $(s_i,m_i)$ angular
momentum variables of the separate electron
and hole, the $(J_i,M_i)$ total angular
momentum variables of the exciton, with
$(J_i=2,M_i=\pm2,\pm1,0)$ or
$(J_i=1,M_i=\pm1,0)$. One can easily go from
one set of operators to the other by
\begin{eqnarray}
B_{\nu_i,\mathbf{Q}_i,J_i,M_i}^\dag&=&\sum_
{s_i,m_i}\langle s_i,m_i|J_i,M_i\rangle\,B_
{\nu_i,\mathbf{Q}_i,s_i,m_i}^\dag\ ,
\\B_{\nu_i,\mathbf{Q}_i,s_i,m_i}^\dag&=&\sum_
{J_i,M_i}\langle J_i,M_i|s_i,m_i\rangle\,B_
{\nu_i,\mathbf{Q}_i,J_i,M_i}^\dag\ .
\end{eqnarray}

Only two of these
$B_{\nu_i,\mathbf{Q}_i,J_i,M_i}^\dag$ states
are coupled to light, namely
$(J_i=1,M_i=\pm1)$, while the six other
exciton states correspond to the so-called
``dark'' excitons. However, even if these dark
excitons are not coupled to light, they are
generated by Coulomb scatterings so that we
must keep these 8 operators
$B_{\nu_i,\mathbf{Q}_i,J_i,M_i}^\dag$ anyway
in order to possibly describe the
exciton-exciton scatterings properly.
Moreover, as the Coulomb interaction is
diagonal within the $(s,m)$ quantum numbers,
but not within the $(J,M)$'s, it is in fact
far simpler to work with the 8 operators
$B_{\nu,\mathbf{Q},s,m}^\dag$ for all
processes dealing with Coulomb interactions,
and just at the beginning and the end of the
calculations, use Eqs (12-13) to transform
these $B_{\nu,\mathbf{Q},s,m}^\dag$ operators
into the $B_{\nu,\mathbf{Q},J,M}^\dag$
operators coupled to the light. This is why
all over this work, we will use the
$B_{\nu,\mathbf{Q},s,m}^\dag$ operators only.

\section{Coulomb creation operator $V_i^\dag$
and direct Coulomb scattering
$\xi_{lnij}^\mathrm{dir}$}

In standard problems with interactions, one
usually divides the system hamiltonian $H$
into $H_0+V$, where $H_0$ is the so-called
``non-interacting'' part, i.\ e.\ the part of
$H$ which can be diagonalized, while $V$ is
the interaction part, i.\ e.\ the part of $H$
which cannot be handled exactly, but is hoped
to be small enough to be treated as a
perturbation. In addition, when $H$ can be
written as $H_0+V$, the $H_0$ eigenstates
form an orthogonal basis which can be used to
expand any state of the system.

For interacting excitons, even if we could
guess that the energies $E_i$ of individual
excitons should appear in the
``non-interacting'' part of an hypothetical
exciton hamiltonian, one cannot divide the
electron-hole Coulomb interaction given in Eq
(5), into a part which would bind a specific
electron to a specific hole to form the $i$
exciton, and a ``rest'' which would make this
exciton to interact with other excitons : The
electrons, as well as the holes, being
indistinguishable particles, such a formal
separation is indeed impossible.

A separation, similar in its spirit to the
splitting $H=H_0+V$, is nevertheless possible
through our commutation technique. If $H$ is
the \emph{exact} semiconductor hamiltonian,
given in Eqs (2-6), and if $B_i^\dag$ is the
\emph{exact} exciton creation operator,
given in Eq (7), we find that their
commutator reads
\begin{equation}
\left[H,B_i^\dag\right]=E_i\,B_i^\dag+V_i^\dag
\ ,
\end{equation}
where $V_i^\dag$ is given by
\begin{eqnarray}
V_i^\dag=\sum_{\mathbf{q}\neq\mathbf{0},\nu_l}
V_\mathbf{q}\,\gamma_{li}(\mathbf{q})\,
B_{\nu_l,\mathbf{Q}_i+\mathbf{q},s_i,m_i}^\dag\,
W_{-\mathbf{q}}\ ,\nonumber
\\ W_{-\mathbf{q}}=\sum_{\mathbf{p},s}
a_{\mathbf{p}-\mathbf{q},s}^\dag\,
a_{\mathbf{p},s}-\sum_{\mathbf{p},m}
b_{\mathbf{p}-\mathbf{q},m}^\dag\,
b_{\mathbf{p},m}\ .
\end{eqnarray}
The coefficient $\gamma_{li}(\mathbf{q})$,
characterizes the scattering of a $\nu_i$
exciton into a $\nu_l$ state under a
$\mathbf{q}$ Coulomb excitation. Using the
following relation,
\begin{equation}
\sum_{\mathbf{k}}\langle x_\nu|\mathbf{k}+
\alpha\mathbf{q}\rangle\,\langle \mathbf{k}|
x_{\nu'}\rangle=\langle
x_\nu|e^{i\alpha\mathbf{q}.\mathbf{r}}|
x_{\nu'}\rangle\ ,
\end{equation}
it is easy to show that it is given by
\begin{equation}
\gamma_{li}(\mathbf{q})=\langle x_{\nu_l}|
e^{i\alpha_h\mathbf{q}.\mathbf{r}}-e^{-i
\alpha_e\mathbf{q}.\mathbf{r}}|x_{\nu_i}
\rangle\ .
\end{equation}

The analogy with the usual separation
$H=H_0+V$ is quite transparent if we note
that Eq (14) also reads
\begin{equation}
H\,B_i^\dag=B_i^\dag\,(H+E_i)+V_i^\dag\ .
\end{equation}
By considering the state
$B_i^\dag|\phi\rangle$, where $|\phi\rangle$
is any electron-hole state, Eq (18) leads to
$$H\,B_i^\dag|\phi\rangle=E_i\,B_i^\dag
|\phi\rangle+B_i^\dag\,H|\phi
\rangle+V_i^\dag|\phi\rangle\ .$$
In the first term, the contribution of the
$i$ exciton to the energy of the system is
just $E_i$ as if this exciton were not
interacting with the other electrons or holes
of $|\phi\rangle$. The second term
corresponds to $H$ acting on $|\phi\rangle$
\emph{independently} of the presence of the
$i$ exciton : The operator $(H+E_i)$, on the
right hand side of $B_i^\dag$, thus plays the
r\^{o}le of the $H_0$ part of the hamiltonian
for usual problems in which $H$ can be
written as $H_0+V$. The third term  is there
because the $i$ exciton does in fact interact
with $|\phi\rangle$. This operator $V_i^\dag$
thus describes all possible Coulomb
interactions between the $i$ exciton and the
rest of the system. It has to be seen as the
formal equivalent of $V$ in $H=H_0+V$. It is
however important to stress that, while the
usual $V$'s conserve particles, i.\ e.\
contain the same number of creation operators
$a^\dag$ and destruction operators $a$, the
operator $V_i^\dag$ is not a real potential
in the sense that it contains one additional
electron-hole pair creation operator $B^\dag$
(see Eq (15)). This is why we call it Coulomb
\emph{creation} operator. In addition, usual
two-body potentials have prefactors which
depend on four indices which are the ones of
two initial and two final states. We here see
that the prefactors appearing in
$V_i^\dag$ depend on two indices only, $l$
and $i$.

In order to cope with these difficulties and
generate a ``scattering'' which depends on
four indices, we can push the commutation
technique one step further and calculate
$\left[ V_i^\dag,B_j^\dag\right]$. Using Eqs
(7-9) and (15), we find
\begin{equation}
\left[ V_i^\dag,B_j^\dag\right]=\sum_{
\mathbf{q}\neq\mathbf{0},\nu_l,\nu_n}
V_{\mathbf{q}}\,\gamma_{li}(\mathbf{q})\,
\gamma_{nj}(-\mathbf{q})\,B_{\nu_l,\mathbf{Q}_i
+\mathbf{q},s_i,m_i}^\dag\,B_{\nu_n,
\mathbf{Q}_j-\mathbf{q},s_j,m_j}^\dag\ ,
\end{equation}
which formally reads
\begin{equation}
\left[ V_i^\dag,B_j^\dag\right]=\sum_{l,n}
\xi_{lnij}^{\mathrm{dir}}\,B_l^\dag\,B_n^\dag\
.
\end{equation}
If we compare Eq (19) to Eq (20), and
symmetrize the result with respect to $(l,n)$
(which will appear convenient afterwards), we
find that the direct Coulomb scattering
$\xi_{lnij}^{\mathrm{dir}}$ can be written as
\begin{equation}
\xi_{lnij}^{\mathrm{dir}}=
\xi_{nlij}^{\mathrm{dir}}=
\xi_{lnji}^{\mathrm{dir}}=\frac{1}{2}\left[
\Delta^{\mathrm{dir}}
\left(^{l\ i}_{n\ j}\right)\hat{\xi}^
{\mathrm{dir}}\left(^{l\ i}_{n\ j}\right) 
+(l\leftrightarrow n)\right]\ .
\end{equation}
It contains an angular momentum contribution
and an orbital contribution which appear as
two independent factors. For direct
processes, this angular momentum
contribution,
\begin{equation}
\Delta^{\mathrm{dir}}
\left(^{l\ i}_{n\ j}\right)=\delta_{s_l,s_i}\,
\delta_{m_l,m_i}\,\delta_{s_n,s_j}\,
\delta_{m_n,m_j}\ ,
\end{equation}
just says that the $l$ exciton has the same
electron and hole momenta as the $i$ exciton
and similarly for the $n$ and $j$ excitons.
Let us stress that this angular momentum part
would be much more complicated if the $(J,M)$
variables for the exciton total angular
momentum were used ; this is why it is indeed
appropriate to keep these $(s,m)$ variables
as long as we deal with Coulomb scatterings
even if they are not the good variables for
semiconductor-photon interaction.

The orbital part of this direct Coulomb
scattering is given by
\begin{equation}
\hat{\xi}^
{\mathrm{dir}}\left(^{l\ i}_{n\ j}\right)=
\delta_{\mathbf{Q}_l+\mathbf{Q}_n,\mathbf{Q}_i
+\mathbf{Q}_j}\,V_{\mathbf{Q}_l-\mathbf{Q}_i}\,
\gamma_{li}(\mathbf{Q}_l-\mathbf{Q}_i)\,
\gamma_{nj}(\mathbf{Q}_n-\mathbf{Q}_j)\ .
\end{equation}
It of course contains the fact that the
center of mass momenta have to be conserved,
$\mathbf{Q}_i+\mathbf{Q}_j=\mathbf{Q}_l+
\mathbf{Q}_n$, in the scattering. It also
contains the two factors which characterize
the scatterings of one exciton from a $\nu_i$
state to a $\nu_l$ state and the other
exciton from a $\nu_j$ state to a $\nu_n$
state, under the Coulomb excitation
$\mathbf{Q}_l-\mathbf{Q}_i=-(\mathbf{Q}_n-
\mathbf{Q}_j)$.

The appendix A contains the explicit
calculation of these
$\gamma_{li}(\mathbf{q})$ factors for the
lowest S states.

Even if the above expression of $\hat{\xi}^
{\mathrm{dir}}\left(^{l\ i}_{n\ j}\right)$ is
quite transparent and extremely convenient
for explicit calculations, it will appear
useful, when we will generate exchange
Coulomb processes, to note that this direct
Coulomb scattering is also equal to
\begin{eqnarray}
\hat{\xi}^
{\mathrm{dir}}\left(^{l\ i}_{n\ j}\right)=
\sum_{\mathbf{k}_l,\mathbf{k}_n,\mathbf{k}_i,
\mathbf{k}_j}\langle x_{\nu_l}|\mathbf{k}_l
\rangle\,\langle x_{\nu_n}|\mathbf{k}_n
\rangle\,\langle \mathbf{k}_i|x_{\nu_i}\rangle
\,\langle \mathbf{k}_j|x_{\nu_j}\rangle
\hspace{5cm}\nonumber
\\ \times\sum_{\mathbf{q}\neq\mathbf{0}}
V_{\mathbf{q}}\left[\delta_{\mathbf{K}_l^e,
\mathbf{K}_i^e+\mathbf{q}}\,\delta_{
\mathbf{K}_l^h,\mathbf{K}_i^h}-(e\leftrightarrow
 h)\right]\left[\delta_{\mathbf{K}_n^e,
\mathbf{K}_j^e-\mathbf{q}}\,\delta_
{\mathbf{K}_n^h,\mathbf{K}_j^h}-(e\leftrightarrow
h)\right]\ .
\end{eqnarray}

In appendix B, we show that this $\hat{\xi}^
{\mathrm{dir}}\left(^{l\ i}_{n\ j}\right)$ is
nothing but the \emph{direct} part of the
Coulomb interaction widely quoted $^{(7,15)}$
in the effective boson exciton hamiltonian,
namely
\begin{eqnarray}
\hat{\xi}^
{\mathrm{dir}}\left(^{l\ i}_{n\ j}\right)=
\int d\mathbf{r}_e\,d\mathbf{r}_h\,
d\mathbf{r}_{e'}\,d\mathbf{r}_{h'}\,\phi_l^\ast
(\mathbf{r}_e,\mathbf{r}_h)\,\phi_n^\ast
(\mathbf{r}_{e'},\mathbf{r}_{h'})\hspace{5cm}
\nonumber
\\ \times\left[V_{ee'}+V_{hh'}-V_{eh'}-V_{e'h}
\right]\,\phi_i(\mathbf{r}_e,\mathbf{r}_h)\,
\phi_j(\mathbf{r}_{e'},\mathbf{r}_{h'})=
\left[\hat{\xi}^
{\mathrm{dir}}\left(^{i\ l}_{j\
n}\right)\right]^\ast\ ,
\end{eqnarray}
where $\phi_i(\mathbf{r}_e,\mathbf{r}_h)$ is
the whole wave function of the $i=(\nu_i,
\mathbf{Q}_i)$ exciton,
\begin{equation}
\langle \mathbf{r}_e,\mathbf{r}_h|B_i^\dag|v
\rangle =\phi_i(\mathbf{r}_e,\mathbf{r}_h)=
\frac{1}{\sqrt{\mathcal{V}}}e^{i\mathbf{Q}_i.
(\alpha_e\mathbf{r}_e+\alpha_h\mathbf{r}_h)}
\langle\mathbf{r}_e-\mathbf{r}_h|x_{\nu_i}
\rangle\ .
\end{equation}

By these two successive commutators
$\left[H,B_i^\dag\right]$ and
$\left[V_i^\dag,B_j^\dag\right]$, we have
found a formal way to generate a scattering of
two excitons $(i,j)$ into two other excitons
$(l,n)$. If we look at its expression given
in Eq (25), we see that the coefficient 
$\hat{\xi}^
{\mathrm{dir}}\left(^{l\ i}_{n\ j}\right)$
obtained by this procedure turns out to have
a very simple physical meaning : It
corresponds to all electron-electron,
hole-hole and electron-hole Coulomb
interactions \emph{between} $(i,j)$ and
$(l,n)$, when the initial and final excitons
are made with the same electron-hole pairs
$(e,h)$ and $(e',h')$.

The scattering processes corresponding to 
$\hat{\xi}^
{\mathrm{dir}}\left(^{l\ i}_{n\ j}\right)$
are shown in Fig.\ (1) : One of the electrons,
or holes, suffers a
$(\mathbf{Q}_l-\mathbf{Q}_i)$ Coulomb
excitation, while another particle has a
$-(\mathbf{Q}_l-\mathbf{Q}_i)$ excitation,
the excitons before and after the scattering
being made with the \emph{same} electron-hole
pairs. This is why we call it a \emph{direct}
Coulomb scattering, by contrast with the
exchange Coulomb scatterings which will
appear below, in which the excitons before
and after scattering are made with different
pairs.

\section{Boson departure operator $D_{ij}$
and bare exchange coefficient
$\lambda_{lnij}$}

Even if the above procedure allows to
\emph{formally} extract from the whole
electron-hole interaction the part which
corresponds to a Coulomb interaction
\emph{between} excitons, such a formal
extraction seeming not obvious at first, we
are far from having picked all the physics
which controls the interactions between
excitons. Another rather subtle origin of
these interactions comes from the fermionic
character of the excitons. From handwaving
arguments, we can say that two excitons feel
each other \emph{even in the absence of any
Coulomb interaction}, because they are made
of two fermions and the fermions of two
excitons must be in different states. This
condition in itself produces a ``link''
between excitons. There is no need of Coulomb
forces. It is thus very likely that the
exciton-exciton interaction has to contain a
purely fermionic contribution, quite
different from the Coulomb interaction
dressed by exchange processes as thought by
everyone up to now.

In this quite tricky determination of the
correct exciton-exciton interaction, it is
however highly necessary to formalize the
above handwaving argument. We want to find a
coefficient, which depends on four exciton
indices $(lnij)$, and which originates from
the fermionic character of the excitons only.
This coefficient has of course to be linked
to the fact that the excitons are not real
bosons. The most direct way to have this
property appearing, is to start with the
commutator of two (exact) exciton creation
operators. It leads to the boson departure
operator $D_{ij}$ defined as
\begin{equation}
D_{ij}=\delta_{ij}-\left[B_i,B_j^\dag\right]\
.
\end{equation}
Using Eq (7) and the standard commutation
rules for the fermion operators $a^\dag$ and
$b^\dag$, we find that this operator reads
\begin{equation}
D_{ij}=\sum_{\mathbf{k}_i,\mathbf{k}_j}
\langle x_{\nu_i}|\mathbf{k}_i\rangle\,
\langle \mathbf{k}_j|x_{\nu_j}\rangle\,
\left[\delta_{m_i,m_j}\,\delta_{\mathbf{K}_i^h,
\mathbf{K}_j^h}\,a_{\mathbf{K}_j^e,s_j}^\dag\,
a_{\mathbf{K}_i^e,s_i}+\delta_{s_i,s_j}\,
\delta_{\mathbf{K}_i^e,\mathbf{K}_j^e}\,
b_{\mathbf{K}_j^h,m_j}^\dag\,b_{\mathbf{K}_i^h,
m_i}\right]\ .
\end{equation}
$D_{ij}\equiv0$ if we have both $s_i\neq s_j$
and $m_i\neq m_j$.

From this boson departure operator, we can
construct a four index coefficient by taking
the commutator of this $D_{ij}$ operator with
an exciton operator. Using Eqs (7,9,28), we
find
\begin{equation}
\left[D_{li},B_j^\dag\right]=2\sum_n\lambda_
{lnij}\,B_n^\dag\ ,
\end{equation}
where the bare exchange coefficient
$\lambda_{lnij}$ can here again be split into
an angular momentum part and an orbital part :
\begin{equation}
\lambda_{lnij}=\lambda_{nlij}=\lambda_{lnji}=
\frac{1}{2}\left[\Delta^{\mathrm{exch}}
\left(^{l\ i}_{n\ j}\right)\,\hat{\lambda}
\left(^{l\ i}_{n\ j}\right)+(l \leftrightarrow
n)\right]\ .
\end{equation}
The angular momentum part,
\begin{equation}
\Delta^{\mathrm{exch}}
\left(^{l\ i}_{n\ j}\right)=\delta_{s_l,s_i}\,
\delta_{m_l,m_j}\,\delta_{s_n,s_j}\,
\delta_{m_n,m_i}\ ,
\end{equation}
just says that the spins of the $l$ and $i$
electrons are the same while the angular
momentum of the $l$ hole is not the one of
the $i$ hole, as in $\Delta^{\mathrm{dir}}
\left(^{l\ i}_{n\ j}\right)$, but the one of
the $j$ hole. The orbital part appears as
\begin{equation}
\hat{\lambda}\left(^{l\ i}_{n\ j}\right)=
\sum_{\mathbf{k}_l,\mathbf{k}_n,\mathbf{k}_i,
\mathbf{k}_j}\langle x_{\nu_l}|\mathbf{k}_l
\rangle\,\langle x_{\nu_n}|\mathbf{k}_n
\rangle\,\langle\mathbf{k}_i|x_{\nu_i}
\rangle\,\langle\mathbf{k}_j|x_{\nu_j}
\rangle\,\delta_{\mathbf{K}_l^e,\mathbf{K}_i^e}
\,\delta_{\mathbf{K}_l^h,\mathbf{K}_j^h}\,
\delta_{\mathbf{K}_n^e,\mathbf{K}_j^e}\,
\delta_{\mathbf{K}_n^h,\mathbf{K}_i^h}\ .
\end{equation}

In appendix B, we show that this orbital part 
$\hat{\lambda}\left(^{l\ i}_{n\ j}\right)$ can
be rewritten in $\mathbf{r}$ space as
\begin{equation}
\hat{\lambda}\left(^{l\ i}_{n\ j}\right)=
\int d\mathbf{r}_e\,d\mathbf{r}_h\,
d\mathbf{r}_{e'}\,d\mathbf{r}_{h'}\,
\phi_l^\ast(\mathbf{r}_e,\mathbf{r}_h)\,
\phi_n^\ast(\mathbf{r}_{e'},\mathbf{r}_{h'})\,
\phi_i(\mathbf{r}_e,\mathbf{r}_{h'})\,
\phi_j(\mathbf{r}_{e'},\mathbf{r}_h)=
\left[\hat{\lambda}\left(^{i\ l}_{j\ n}\right)
\right]^\ast\ .
\end{equation}
From the above expression, we see that, in
this orbital part, the electrons and the
holes forming the excitons are exchanged, the
$l$ and $i$ excitons being made with the same
electron but different holes. Note that the
$(l\leftrightarrow n)$ term of Eq (30)
restores the $(e,h)$ symmetry in
$\lambda_{lnij}$, which is broken in 
$\hat{\lambda}\left(^{l\ i}_{n\ j}\right)$.

Eq (33) shows in a transparent way that this
$\hat{\lambda}\left(^{l\ i}_{n\ j}\right)$
exchange coefficient can exist because of the
composite nature of the excitons which can be
formed in different ways, the electrons and
the holes being indistinguishable fermions.
An equally transparent link between this
coefficient and the composite nature of the
excitons can be obtained by considering the
two-exciton operator $B_i^\dag\,B_j^\dag$,
each $B^\dag$ being given by Eq (7), and by
binding the electrons and holes of these two
$B^\dag$'s in a different way through Eq (9).
This gives
\begin{equation}
B_i^\dag\,B_j^\dag=-\sum_{l,n}\lambda_{lnij}
\,B_l^\dag\,B_n^\dag\ ,
\end{equation}
with the same prefactor $\lambda_{lnij}$ as
the one of Eq (29).

The electron-hole exchange appearing in the
first term of Eq (30) is shown in Fig.\ (2).
It corresponds to cross the holes when
forming the $(i,j)$ or $(l,n)$ excitons : It
is indeed a bare exchange process in the
sense that it does not contain any Coulomb
interaction, by contrast with the Coulomb
exchange processes which will appear later on.

The four $\delta$'s appearing in
$\hat{\lambda}\left(^{l\ i}_{n\ j}\right)$
given in Eq (32) impose the expected exciton
total momentum conservation $\mathbf{Q}_i+
\mathbf{Q}_j=\mathbf{Q}_l+\mathbf{Q}_n$ as
well as $\mathbf{k}_i+\mathbf{k}_j=
\mathbf{k}_l+\mathbf{k}_n$. By expliciting
these four $\delta$'s, we can rewrite
$\hat{\lambda}\left(^{l\ i}_{n\ j}\right)$ in
a compact form,
\begin{equation}
\hat{\lambda}\left(^{l\ i}_{n\ j}\right)=
\delta_{\mathbf{Q}_l+\mathbf{Q}_n,\mathbf{Q}_i
+\mathbf{Q}_j}\,F_{lnij}\left(\alpha_e
(\mathbf{Q}_l-\mathbf{Q}_i),\alpha_h
(\mathbf{Q}_n-\mathbf{Q}_i)\right)\ ,
\end{equation}
where $F_{lnij}(\mathbf{p},\mathbf{p'})$ is a
sum over one momentum only,
\begin{equation}
F_{lnij}(\mathbf{p},\mathbf{p'})=
\sum_{\mathbf{k}}\langle
x_{\nu_l}|\mathbf{k}-\frac{\mathbf{p}+\mathbf{p'}}
{2}\rangle\,\langle x_{\nu_n}|\mathbf{k}+
\frac{\mathbf{p}+\mathbf{p'}}{2}\rangle\,
\langle\mathbf{k}+\frac{\mathbf{p}-\mathbf{p'}}
{2}|x_{\nu_i}\rangle\,\langle\mathbf{k}-
\frac{\mathbf{p}-\mathbf{p'}}{2}|x_{\nu_j}
\rangle\ .
\end{equation}
Although less transparent for physical
understanding, the above expression of
$\hat{\lambda}\left(^{l\ i}_{n\ j}\right)$ is
more convenient for explicit calculations.
When the four excitons are in S states, the
function $F_{lnij}(\mathbf{p},\mathbf{p'})$ a
priori depends on three independent
parameters, $p$, $p'$ and the angle $\theta$
between $\mathbf{p}$ and $\mathbf{p'}$. Some
interesting values of these parameters are :

(i) $p=0$ or $p'=0$, which corresponds to
$\mathbf{Q}_l=\mathbf{Q}_i$ or
$\mathbf{Q}_n=\mathbf{Q}_i$ : The final
excitons have the same momenta as the initial
excitons ;

(ii) $\theta=0$ or $\theta=\pi$, which
corresponds to $\mathbf{p}$ parallel to
$\mathbf{p'}$.
$\mathbf{Q}_l-\mathbf{Q}_i$ is parallel to
$\mathbf{Q}_n-\mathbf{Q}_i\equiv
\mathbf{Q}_j-\mathbf{Q}_l$ in the particular
case of $\mathbf{Q}_i=\mathbf{Q}_j$ or
$\mathbf{Q}_l=\mathbf{Q}_n$, i.\ e.\ when the
two initial or the two final excitons have
the same momentum. In this case, the two
independent parameters can be taken either as
$\alpha_e|\mathbf{Q}_l-\mathbf{Q}_i|$ and
$\alpha_h|\mathbf{Q}_n-\mathbf{Q}_i|$, or
better as the momentum transfer 
$|\mathbf{Q}_l-\mathbf{Q}_i|$ and the mass
ratio $m_e/m_h$.

In appendix C, we have calculated
$F_{lnij}(\mathbf{p},\mathbf{p'})$ when all
the $\nu$'s are equal to the 1S ground state
and $\mathbf{Q}_i=\mathbf{Q}_j=\mathbf{Q}_l=
\mathbf{Q}_n$. We get
\begin{eqnarray}
F_{1s1s1s1s}(\mathbf{0},\mathbf{0})&=&
(33\pi/2)a_x^3/\mathcal{V}\hspace{1,5
cm}\mathrm{in}\ 3\mathrm{D}\nonumber
\\ &=& (4\pi/5)a_x^2/\mathcal{S}\hspace{1.7
cm}\mathrm{in}\ 2\mathrm{D}\ ,
\end{eqnarray}
$a_x$ being the 3D exciton Bohr radius.

We have also numerically computed
$\hat{\lambda}\left(^{l\ i}_{n\ j}\right)$
in the 2D case, when
$\nu_l=\nu_n=\nu_i=\nu_j=1\mathrm{S}$,
$\mathbf{Q}_i=\mathbf{Q}_j=\mathbf{0}$,
$\mathbf{Q}_l=-\mathbf{Q}_n=\mathbf{p}$.
Fig.\ (3) shows
\begin{equation}
\Lambda(p)=\frac{\mathcal{S}}{\lambda_{2D}^2}
\hat{\lambda}\left(_{-\mathbf{p}\ \mathbf{0}}
^{\ \,\mathbf{p}\ \,\mathbf{0}}\right)\ ,
\end{equation}
for three values of $m_e/m_h$, as a function
of $P=p\lambda_{2D}$, with
$\lambda_{2D}=a_x/2$ being the 2D Bohr
radius. We see that $\Lambda(p)$ weakly
depends on the mass ratio, and that it is
significant for values of $p$ such that
$p\lambda_{2D}<4$ only.

\section{Coulomb exchange scatterings}

From Eq (33), we see that the
$\lambda_{lnij}$ coefficient is
dimensionless. So that it has to be
``cooked'' with quantities homogeneous to an
energy in order to possibly appear in an
exciton-exciton scattering. From the two
energy-like quantities we have yet found in
this problem, namely the bare energies of
the excitons $E_i$ and the direct Coulomb
scattering $\xi_{lnij}^{\mathrm{dir}}$,
there are of course various ways to
construct a $\lambda_{lnij}$ dependent
scattering. In the next section, we will
show that two specific combinations of the
bare exchange coefficient $\lambda_{lnij}$
and the direct Coulomb scattering
$\xi_{lnij}^{\mathrm{dir}}$ appear in a
natural way, namely
\begin{equation}
\xi_{lnij}^{\mathrm{right}}=\sum_{p,r}
\xi_{lnpr}^\mathrm{dir}\,\lambda_{prij}\ ,
\end{equation} 
\begin{equation}
\xi_{lnij}^{\mathrm{left}}=\sum_{p,r}
\lambda_{lnpr}\,\xi_{prij}^\mathrm{dir}\ .
\end{equation}
They are shown in Fig.\ (4).

The simplest way to calculate these sums is
to use Eqs (30-32) for $\lambda_{lnij}$ and
Eqs (21,22,24) for
$\xi_{lnij}^{\mathrm{dir}}$. We find
\begin{equation}
\xi_{lnij}^{\mathrm{right}}=\frac{1}{2}
\left(\Delta^{\mathrm{exch}}\left(_{n\ j}
^{l\ i}\right)\,\hat{\xi}^{\mathrm{right}}
\left(_{n\ j}^{l\ i}\right)+(l\leftrightarrow
n)\right)\ ,
\end{equation}
where the angular momentum conservation part
$\Delta^{\mathrm{exch}}\left(_{n\ j}
^{l\ i}\right)$ is just the one appearing in
the bare exchange coefficient
$\lambda_{lnij}$ (Eq (31)), while the orbital
part of this right exchange Coulomb term is
given by
\begin{eqnarray}
\hat{\xi}^{\mathrm{right}}
\left(_{n\ j}^{l\ i}\right)=\sum_{p,r}
\hat{\xi}^{\mathrm{dir}}
\left(_{n\ r}^{l\ p}\right)
\hat{\lambda}
\left(_{r\ j}^{p\
i}\right)\hspace{9cm}\nonumber
\\=\sum_{\mathbf{k}_l,\mathbf{k}_n,\mathbf{k}
_i,\mathbf{k}_j}\langle x_{\nu_l}|\mathbf{k}_l
\rangle\,\langle x_{\nu_n}|\mathbf{k}_n
\rangle\,\langle\mathbf{k}_i|x_{\nu_i}
\rangle\,\langle\mathbf{k}_j|x_{\nu_j}
\rangle\hspace{5.2 cm}\nonumber
\\ \times\sum_{\mathbf{q}\neq\mathbf{0}}
V_{\mathbf{q}}
\left[\delta_{\mathbf{K}_l^e,
\mathbf{K}_i^e+\mathbf{q}}\delta_{\mathbf{K}_l^h,
\mathbf{K}_j^h}-\delta_{\mathbf{K}_l^h,
\mathbf{K}_j^h+\mathbf{q}}\delta_{\mathbf{K}_l^e,
\mathbf{K}_i^e}\right]\left[
\delta_{\mathbf{K}_n^e,\mathbf{K}_j^e-
\mathbf{q}}
\delta_{\mathbf{K}_n^h,\mathbf{K}_i^h}-
\delta_{\mathbf{K}_n^h,\mathbf{K}_i^h-
\mathbf{q}}
\delta_{\mathbf{K}_n^e,\mathbf{K}_j^e}\right]\
.
\end{eqnarray}
Note that the second term of each bracket
does \emph{not} correspond to
$(e\leftrightarrow h)$ as for
$\hat{\xi}^{\mathrm{dir}}\left(_{n\ j}^{l\
i}\right)$ given in Eq (24).

The set of four $\delta$'s appearing in 
$\hat{\xi}^{\mathrm{right}}\left(_{n\ j}^{l\
i}\right)$ again impose the expected momentum
conservation $\mathbf{Q}_i+\mathbf{Q}_j=
\mathbf{Q}_l+\mathbf{Q}_n$. By expliciting
these four $\delta$ functions, it is possible
to rewrite $\hat{\xi}^{\mathrm{right}}
\left(_{n\ j}^{l\ i}\right)$ in a more
compact form in terms of sums over
$\mathbf{q}$ and one $\mathbf{k}$ only. In
addition to the function $F_{lnij}(\mathbf{p},
\mathbf{p'})$ appearing already in the bare
exchange coefficient (Eq (36)), two other
functions enter  $\hat{\xi}^{\mathrm{right}}
\left(_{n\ j}^{l\ i}\right)$. They are
defined by
\begin{equation}
G_{lnij}^{(i)}(\mathbf{p},\mathbf{p'};
\mathbf{q})=\sum_{\mathbf{k}}
\langle x_{\nu_l}|\mathbf{k}-\frac{\mathbf{p}
+\mathbf{p'}}{2}\rangle\,\langle x_{\nu_n}|
\mathbf{k}+\frac{\mathbf{p}+\mathbf{p'}}{2}
\rangle\,\langle\mathbf{k}+\frac{\mathbf{p}-
\mathbf{p'}}{2}+\mathbf{q}|x_{\nu_i}\rangle\,
\langle\mathbf{k}-\frac{\mathbf{p}-\mathbf{p'}}
{2}|x_{\nu_j}\rangle\ ,
\end{equation}
\begin{equation}
G_{lnij}^{(j)}(\mathbf{p},\mathbf{p'};
\mathbf{q})=\sum_{\mathbf{k}}
\langle x_{\nu_l}|\mathbf{k}-\frac{\mathbf{p}
+\mathbf{p'}}{2}\rangle\,\langle x_{\nu_n}|
\mathbf{k}+\frac{\mathbf{p}+\mathbf{p'}}{2}
\rangle\,\langle\mathbf{k}+\frac{\mathbf{p}-
\mathbf{p'}}{2}|x_{\nu_i}\rangle\,
\langle\mathbf{k}-\frac{\mathbf{p}-\mathbf{p'}}
{2}+\mathbf{q}|x_{\nu_j}\rangle\ .
\end{equation}
(We can see that
$F_{lnij}(\mathbf{p},\mathbf{p'})$ is also
either $G_{lnij}^{(i)}(\mathbf{p},\mathbf{p'};
\mathbf{0})$ or
$G_{lnij}^{(j)}(\mathbf{p},\mathbf{p'};
\mathbf{0})$). In terms of these three
functions, the right exchange Coulomb term
reads
\begin{eqnarray}
\hat{\xi}^{\mathrm{right}}
\left(_{n\ j}^{l\ i}\right)=\delta_
{\mathbf{Q}_l+\mathbf{Q}_n,\mathbf{Q}_i+
\mathbf{Q}_j}\sum_{\mathbf{q}\neq\mathbf{0}}
V_{\mathbf{q}}\hspace{9.5cm}\nonumber
\\ \times\left[F_{lnij}\left(\alpha_e
(\mathbf{Q}_l-\mathbf{Q}_i)-\mathbf{q},
\alpha_h(\mathbf{Q}_n-\mathbf{Q}_i)\right)+
F_{lnij}\left(\alpha_e(\mathbf{Q}_l-\mathbf{Q}_i)
,\alpha_h(\mathbf{Q}_n-\mathbf{Q}_i)+
\mathbf{q}\right)\right.\nonumber
\\ \left. -G_{lnij}^{(i)}\left(\alpha_e(
\mathbf{Q}_l-\mathbf{Q}_i),\alpha_h
(\mathbf{Q}_n-\mathbf{Q}_i);-\mathbf{q}\right)
-G_{lnij}^{(j)}\left(\alpha_e(\mathbf{Q}_l-
\mathbf{Q}_i),\alpha_h(\mathbf{Q}_n-\mathbf{Q}_i)
;\mathbf{q}\right)\right].
\end{eqnarray}

This $\hat{\xi}^{\mathrm{right}}
\left(_{n\ j}^{l\ i}\right)$ coefficient can
be rewritten in a form much more transparent
for the physical understanding, although much
less convenient for calculations. In appendix
B, we do show that
\begin{eqnarray}
\hat{\xi}^{\mathrm{right}}
\left(_{n\ j}^{l\ i}\right)=\int
d\mathbf{r}_e\,d\mathbf{r}_h\,d\mathbf{r}_{e'}\,
d\mathbf{r}_{h'}\,\phi_l^\ast(\mathbf{r}_e,
\mathbf{r}_h)\,\phi_n^\ast(\mathbf{r}_{e'},
\mathbf{r}_{h'})\hspace{3cm}\nonumber
\\ \times\left[V_{ee'}+V_{hh'}-V_{eh'}-V_{e'h}
\right]\phi_i(\mathbf{r}_e,\mathbf{r}_{h'})\,
\phi_j(\mathbf{r}_{e'},\mathbf{r}_h)\ .
\end{eqnarray}
The physical meaning of this $\hat{\xi}^
{\mathrm{right}}
\left(_{n\ j}^{l\ i}\right)$ becomes now
clear : It contains the electron-electron and
hole-hole interactions \emph{between} two
excitons when these excitons are made with
their electrons and holes coupled in a
different way. With respect to the
electron-hole interactions, the situation is
however more subtle as these interactions are
\emph{between} the excitons on one side, but
\emph{inside} the excitons of the other side,
due to the exchange of the electrons or holes
making the excitons of the two sides. This
$\hat{\xi}^ {\mathrm{right}}
\left(_{n\ j}^{l\ i}\right)$ turns out to be
nothing but the exchange Coulomb interaction
which appears in the exciton-exciton
scattering coefficient of the effective boson
exciton hamiltonian widely quoted $^{(7)}$.

The coefficient $\xi_{lnij}^{\mathrm{left}}$
reads as $\xi_{lnij}^{\mathrm{right}}$ except
that $V_{eh'}+V_{e'h}$ is replaced by
$V_{eh}+V_{e'h'}$ in Eq (46), so that the
electron-hole Coulomb interactions are now
\emph{between} the $(i,j)$ excitons and
\emph{inside} the $(l,n)$ excitons. As a
consequence, the two coefficients
$\xi^{\mathrm{left}}$ and
$\xi^{\mathrm{right}}$ verify
\begin{equation}
\xi_{lnij}^{\mathrm{left}}=\left[
\xi_{ijln}^{\mathrm{right}}\right]^\ast\ ,
\end{equation}
with a similar relation between the orbital
parts $\hat{\xi}^{\mathrm{left}}$ and 
$\hat{\xi}^{\mathrm{right}}$. This can also
be seen from Eqs (25,33,39,40).

From $\xi_{lnij}^{\mathrm{right}}$ and
$\xi_{lnij}^{\mathrm{left}}$, it is possible
to construct a quantity symmetrical with
respect to the electron-hole interactions, by
taking
\begin{equation}
\xi_{lnij}^{\mathrm{exch}}=\frac{1}{2}\left(
\xi_{lnij}^{\mathrm{right}}+
\xi_{lnij}^{\mathrm{left}}\right)\ .
\end{equation}
Using Eq (46), the corresponding orbital part
of this Coulomb exchange coefficient reads
\begin{eqnarray}
\hat{\xi}^{\mathrm{exch}}\left(_{n\ j}^{l\
i}\right)=\int d\mathbf{r}_e\,d\mathbf{r}_h\,
d\mathbf{r}_{e'}\,d\mathbf{r}_{h'}\,
\phi_l^\ast(\mathbf{r}_e,\mathbf{r}_h)\,
\phi_n^\ast(\mathbf{r}_{e'},\mathbf{r}_{h'})
\hspace{5.5 cm}\nonumber
\\ \times\left[V_{ee'}+V_{hh'}-\frac{1}{2}(
V_{eh}+V_{e'h'}+V_{eh'}+V_{e'h}\right]
\phi_i(\mathbf{r}_e,\mathbf{r}_{h'})\,
\phi_j(\mathbf{r}_{e'},\mathbf{r}_h)\ .
\end{eqnarray}
It contains the four possible electron-hole
Coulomb interactions between two electrons
and two holes. This
$\xi_{lnij}^{\mathrm{exch}}$ could be a
reasonable scattering for the interaction
part of an hypothetical effective exciton
hamiltonian : As $\xi_{lnij}^{\mathrm{exch}}=
\left(\xi_{ijln}^{\mathrm{exch}}\right)^\ast$,
its contribution would be hermitian, which is
not the case neither for
$\xi_{lnij}^{\mathrm{right}}$ nor for
$\xi_{lnij}^{\mathrm{left}}$ alone.

\section{Matrix element of $H$ between
two-exciton states}

In some physical problems such as the one
considered in section VI of this paper, we
are led to consider the matrix elements of
the exact hamiltonian $H$ between two-exciton
states, namely $\langle
v|B_lB_nHB_i^\dag B_j^\dag|v\rangle$. Using
Eqs (14,20) and $V_i^\dag|v\rangle=0$, as
obvious from Eq (15), we find
\begin{equation}
H\,B_i^\dag\,B_j^\dag|v\rangle=(E_i+E_j)
B_i^\dag\,B_j^\dag|v\rangle+\sum_{p,r}
\xi_{prij}^{\mathrm{dir}}\,B_p^\dag\,
B_r^\dag|v\rangle\ .
\end{equation}
So that this matrix element reads
\begin{equation}
\langle v|B_lB_nHB_i^\dag B_j^\dag|v\rangle=
(E_i+E_j)\langle v|B_lB_nB_i^\dag B_j^\dag|v
\rangle+\sum_{p,r}\xi_{prij}^{\mathrm{dir}}
\langle v|B_lB_nB_p^\dag B_r^\dag|v\rangle\ .
\end{equation}

To go further, we must calculate $\langle 
v|B_lB_nB_i^\dag B_j^\dag|v\rangle$. This is
easily done using Eqs (27) and (29). Since
$D_{ij}|v\rangle=0$, as obvious from Eq (28),
we get
\begin{equation}
\langle 
v|B_lB_nB_i^\dag B_j^\dag|v\rangle=
\delta_{li}\,\delta_{nj}+\delta_{ni}\,\delta
_{lj}-2\lambda_{lnij}\ .
\end{equation}
The two first terms are na\"{\i}ve : They
differ from zero for $(l,n)=(i,j)$, i.\ e.\
when the excitons on both sides are
identical. The last term is more subtle. It
has a fermionic origin : In addition to the
fact that the coefficient $\lambda_{lnij}$ is
directly related to the boson deviation
operator $D_{li}$ through Eq (29), its
fermionic origin can also be traced back to
Eq (34) which says that a product of two
exciton operators writes as a sum of products
of any two other $B^\dag$'s so that $B_i^\dag
B_j^\dag$ in fact contains a ``piece'' of any
$B_l^\dag B_n^\dag$, even if
$(l,n)\neq(i,j)$. In other words, the
$B_i^\dag B_j^\dag|v\rangle$ states do not
form an orthogonal basis for two-pair states.

By inserting Eq (52) into Eq (51), we get
\begin{equation}
\langle v|B_lB_nHB_i^\dag B_j^\dag|v\rangle=
(E_i+E_j)(\delta_{li}\,\delta_{nj}+
\delta_{lj}\,\delta_{ni}-2\lambda_{lnij})
+2(\xi_{lnij}^{\mathrm{dir}}-
\xi_{lnij}^{\mathrm{left}})\ .
\end{equation}
This calculation thus produces the Coulomb
exchange term $\xi_{lnij}^{\mathrm{left}}$. A
similar calculation done with $H$ acting on
the left gives
\begin{equation}
\langle v|B_lB_nHB_i^\dag B_j^\dag|v\rangle=
(E_l+E_n)(\delta_{li}\,\delta_{nj}+
\delta_{lj}\,\delta_{ni}-2\lambda_{lnij})
+2(\xi_{lnij}^{\mathrm{dir}}-
\xi_{lnij}^{\mathrm{right}})\ .
\end{equation}
Here appears $\xi_{lnij}^{\mathrm{right}}$.
From these two expressions of the $H$ matrix
element, we deduce that 
$\xi_{lnij}^{\mathrm{right}}$ and
$\xi_{lnij}^{\mathrm{left}}$ are linked by
\begin{equation}
\xi_{lnij}^{\mathrm{left}}-
\xi_{lnij}^{\mathrm{right}}=(E_l+E_n-E_i-E_j)
\lambda_{lnij}\ ,
\end{equation}
so that they are equal for $E_l+E_n=E_i+E_j$
only.

It is possible to check that Eq (55) remains
valid when $\xi$ and $\lambda$ are replaced
by their orbital parts $\hat{\xi}$ and
$\hat{\lambda}$. Using this equation, we can
easily calculate the difference
$\hat{\xi}^{\mathrm{left}}-
\hat{\xi}^{\mathrm{right}}$ and compare it to
$\hat{\xi}^{\mathrm{left}}$ (or
$\hat{\xi}^{\mathrm{right}}$). Ciuti et al
$^{(11)}$ and Rochat et al $^{(9)}$ have
calculated
$\hat{\xi}^{\mathrm{left}}\left(_{n\ j}^{l\
i}\right)$ for 2D scatterings between
$\nu_i=\nu_j=1\mathrm{S}$,
$\mathbf{Q}_i=\mathbf{Q}_j=\mathbf{0}$, and
$\nu_l=\nu_n=1\mathrm{S}$,
$\mathbf{Q}_l=-\mathbf{Q}_n=\mathbf{p}$.
Fig.\ (4) of reference (9) precisely shows
\begin{equation}
g^{\mathrm{exch}}(p)=-\frac{\pi^2\mathcal{S}}
{4e^2\lambda_{2D}}\hat{\xi}^{\mathrm{left}}\left(_{-\mathbf{p}\ \mathbf{0}}
^{\ \,\mathbf{p}\ \,\mathbf{0}}\right)
\ ,
\end{equation}
as a function of $P=p\lambda_{2D}$, with
$\lambda_{2D}$ being the 2D Bohr radius. On
our Fig.\ (5) is shown
\begin{equation}
\Delta
g(p)=\frac{\pi^2\mathcal{S}}{4e^2\lambda_{2D}}
\left[\hat{\xi}^{\mathrm{left}}\left(_{-\mathbf{p}\ \mathbf{0}}
^{\ \,\mathbf{p}\ \,\mathbf{0}}\right)
-
\hat{\xi}^{\mathrm{right}}\left(_{-\mathbf{p}\ \mathbf{0}}
^{\ \,\mathbf{p}\ \,\mathbf{0}}\right)
\right]=
\frac{\pi^2}{2}\alpha_e\alpha_h
p^2\lambda_{2D}^2 \Lambda(p)\ ,
\end{equation}
with $\Lambda(p)$ given by Eq (38). We see
that, except for very small $p\lambda_{2D}$,
this difference is of the order of
$g^{\mathrm{exch}}(p)$, unless $m_e<<m_h$.

We can rewrite the $H$ matrix element between
two-exciton states in a more symmetrical form
with respect to $(l,n)$ and $(i,j)$ by taking
half the sum of Eqs (53) and (54). If in
addition, we consider this $H$ matrix element
between two-exciton states \emph{normalized},
but still not orthogonal, namely
\begin{equation}
|\psi_{ji}\rangle=|\psi_{ij}\rangle=\frac
{B_i^\dag B_j^\dag|v\rangle}
{\langle v|B_iB_jB_j^\dag B_i^\dag|v\rangle
^{1/2}}=\frac{1}{\mathcal{N}_{ij}}B_i^\dag
B_j^\dag|v\rangle\ ,
\end{equation}
where $\mathcal{N}_{ij}=(1+\delta_{ij}-2
\lambda_{ijij})^{1/2}$ due to Eq (52), we
find, using Eqs (53-54),
\begin{equation}
\langle\psi_{ln}|H|\psi_{ij}\rangle=\frac{1}{2}
(E_l+E_n+E_i+E_j)\,\delta_{lnij}+\frac{2}
{\mathcal{N}_{ln}\mathcal{N}_{ij}}\left[
\xi_{lnij}^{\mathrm{dir}}-
\xi_{lnij}^{\mathrm{exch}}-
(1-\delta_{lnij})\eta_{lnij}\right]\ .
\end{equation}
$\delta_{lnij}$ is equal to 1 when the
excitons on both sides are identical i.\ e.\
$(l,n)=(i,j)$, and 0 otherwise, while
$\eta_{lnij}$ is given by
\begin{equation}
\eta_{lnij}=\frac{1}{2}(E_l+E_n+E_i+E_j)
\lambda_{lnij}\ .
\end{equation}

The first term of this $H$ matrix element
just corresponds to the energies of the
\emph{non-interacting} $(i,j)$ or $(l,n)$
excitons if, \emph{in addition}, the
two-exciton states $|\psi_{ij}\rangle$ are
assumed to be orthogonal. The last terms of
this $H$ matrix element come from
interactions. They have three origins. The
first term,
$\xi_{lnij}^{\mathrm{dir}}$, is a direct
Coulomb scattering between excitons made on
both sides with the same electrons and
holes. The second term,
$\xi_{lnij}^{\mathrm{exch}}$, is an exchange
Coulomb scattering in which the holes (or
electrons) making the excitons are exchanged.
The last term, $\eta_{lnij}$, has a purely
fermionic origin : It is directly linked to
the fact that the $|\psi_{ij}\rangle$ and
$|\psi_{ln}\rangle$ states are not orthogonal
for $(l,n)\neq(i,j)$. It does not exist in
diagonal scatterings. Let us however stress
that, as $\eta_{lnij}$ depends on the
\emph{sum} of the four exciton energies, it
is band gap dependent ; so that it is very
unlikely that such a $\eta_{lnij}$ term
appears in a physical scattering. We are
going to come back to this problem in the
next paragraph.

It might be useful to mention that the
diagonal matrix element of $H$ reduces to
\begin{equation}
\langle\psi_{ij}|H|\psi_{ij}\rangle=E_i+E_j
+(\xi_{ijij}^{\mathrm{dir}}-
\xi_{ijij}^{\mathrm{exch}})/(1+\delta_{ij}-2
\lambda_{ijij})\ ,
\end{equation}
with $\xi_{ijij}^{\mathrm{exch}}$ possibly
replaced by $\xi_{ijij}^{\mathrm{right}}$ or
$\xi_{ijij}^{\mathrm{left}}$ as these three
scatterings are equal for diagonal processes.
$(E_i+E_j)$ is just what we would expect for
the expectation value of the hamiltonian
between two \emph{non-interacting}
boson-excitons
$i$ and $j$. The other terms come from
scattering processes between the $i$ and $j$
excitons resulting from Coulomb interaction
and Pauli exclusion.

\section{Exciton-exciton scattering rate}

Let us end this work by considering a
physical quantity in which these $H$ matrix
elements between two-exciton states may appear
in a direct way. Ciuti et al (CSPQS) have
proposed \nolinebreak
$^{(11)}$ to calculate the elastic Coulomb
scattering of 1s excitons by 1s excitons,
namely
\begin{equation}
(1\mathrm{s},\mathbf{Q},S)+(1\mathrm{s},
\mathbf{Q'},S')\ \rightarrow\
(1\mathrm{s},\mathbf{Q}+\mathbf{q},S_f)+
(1\mathrm{s},\mathbf{Q'}-\mathbf{q},S'_f)\ ,
\end{equation}
(see their Eq (3)), from the matrix element
of the exact hamiltonian $H$ between these
two-exciton states. Let us
rewrite this Eq (62) as
$(i)+(j)\rightarrow(l)+(n)$. In the following
we will drop the angular momentum parts for
simplicity, as they are unimportant for the
problem we raise.

The one-exciton wave function
$\phi_{\mathbf{Q}}(\mathbf{r}_e,\mathbf{r}_h)$
given in their Eq (1) is just our
$\phi_i(\mathbf{r}_e,\mathbf{r}_h)$ (see Eq
(26)), while the two-exciton wave function
$\phi_{\mathbf{QQ'}}^{SS'}$ given in their Eq
(4) reads, within our notations,
\begin{equation}
\phi_{ij}(\mathbf{r}_e,\mathbf{r}_h,\mathbf{r}
_{e'},\mathbf{r}_{h'})=\frac{1}{2}\left[
\phi_i(\mathbf{r}_e,\mathbf{r}_h)\,\phi_j
(\mathbf{r}_{e'},\mathbf{r}_{h'})
+(\mathbf{r}_e,\mathbf{r}_h\leftrightarrow
\mathbf{r}_{e'},\mathbf{r}_{h'})-
(\mathbf{r}_e\leftrightarrow\mathbf{r}_{e'})
-(\mathbf{r}_h\leftrightarrow\mathbf{r}_{h'})
\right]\ ,
\end{equation}
so that it is just the wave function of the
state $B_i^\dag B_j^\dag|v\rangle$.

It is possible to show directly from Eq (63),
i.\ e.\ their Eq (4), that
\begin{equation}
\langle\phi_{ln}|\phi_{ij}\rangle=
\delta_{li}\,\delta_{nj}+\delta_{lj}\,
\delta_{ni}-2\lambda_{lnij}\ ,
\end{equation}
with
\begin{equation}
\lambda_{lnij}=\frac{1}{2}\int d\mathbf{r}_e\,
d\mathbf{r}_h\,d\mathbf{r}_{e'}\,
d\mathbf{r}_{h'}\left[\phi_l^\ast(\mathbf{r}_e,
\mathbf{r}_h)\,\phi_n^\ast(\mathbf{r}_{e'},
\mathbf{r}_{h'})+(l\leftrightarrow n)\right]
\phi_i(\mathbf{r}_e,\mathbf{r}_{h'})\,
\phi_j(\mathbf{r}_{e'},\mathbf{r}_h)\ ,
\end{equation}
which is nothing but our Eqs (30,33) and
(52). As a consequence, the
$|\phi_{ln}\rangle$ states are \emph{not}
normalized. Eq (64) also shows that the
initial and final states of the
exciton-exciton scattering considered in Eq
(62) are \emph{not} orthogonal for
$\mathbf{q}\neq\mathbf{0}$ or
$\mathbf{q}\neq\mathbf{Q'}-\mathbf{Q}$. So
that, even if for such $\mathbf{q}$, the
$\delta$ terms of Eq (64) give zero, this
matrix element is not zero due to the
$\lambda_{lnij}$ term.

According to CSPQS, the scattering amplitude
corresponding to the process of Eq (62)
should be equal to
$H_{SS'}^{S_fS'_f}(\mathbf{Q},\mathbf{Q'},
\mathbf{q})=\langle\phi_{ij}|H|\phi_{ln}
\rangle$ (see their Eq (6)), where $H$ is the
exact two-electron and two-hole semiconductor
hamiltonian. Their Eq (6) is nothing but
$\langle v|B_iB_jHB_l^\dag B_n^\dag|v\rangle$
written in $\mathbf{r}$ space. Our equations
(53,54) immediately give
\begin{eqnarray}
\langle v|B_iB_jHB_l^\dag B_n^\dag|v\rangle
&=&2\left[\xi_{ijln}^{\mathrm{dir}}-
\xi_{ijln}^{\mathrm{left}}-(E_l+E_n)\lambda_
{ijln}\right]
\\ &=&2\left[\xi_{ijln}^{\mathrm{dir}}-
\xi_{ijln}^{\mathrm{right}}-(E_i+E_j)\lambda_
{ijln}\right]\ ,
\end{eqnarray}
as the $\delta$ terms give zero for
$\mathbf{q}\neq\mathbf{0}$ and $\mathbf{q}\neq
\mathbf{Q'}-\mathbf{Q}$. If we consider the
results given by CSPQS in their Eq (7), we see
that the first two terms given by their Eqs
(12-14), exactly correspond to the two terms
of $(2\xi_{ijln}^{\mathrm{dir}})$ (see our
Eqs (21,25)), while the last two terms
(their Eq (16) and its ``hole'' equivalent)
correspond to the two terms of
$(-2\xi_{ijln}^{\mathrm{right}})$ (see our
Eqs (41,46)). Consequently they should have
found the result given in Eq (67). Even if
calculations in $\mathbf{r}$ space are quite
cumbersome, it is in fact possible to check
the existence of the missing term $(E_i+E_j)
\lambda_{ijln}$ directly from a (tedious)
calculation of the right hand side of their
Eq (6). The origin of their mistake probably
comes from the fact that they have considered
the two states
$|\phi_{ln}\rangle$ and $|\phi_{ij}\rangle$
as orthogonal, which is not
true (see Eq (64)).

The correct value of the
$\langle\phi_{ij}|H|\phi_{ln}\rangle$ matrix
element given in Eqs (66-67) leads us to
question the validity of the whole procedure
to determine the exciton-exciton scattering
rate.

(i) First the $|\phi_{ij}\rangle$ wave
functions are not normalized, so that it
would be reasonable to use normalized states
and relate the exciton-exciton scattering not
to $\langle\phi_{ij}|H|\phi_{ln}\rangle$ but
to $\langle\phi_{ij}|H|\phi_{ln}\rangle/
\langle\phi_{ij}|\phi_{ij}\rangle^{1/2}
\langle\phi_{ln}|\phi_{ln}\rangle^{1/2}$. This
would add a prefactor to all scatterings
(see Eqs (58-59)).

(ii) There is however a much more dramatic
problem with CSPQS procedure. The existence of
the last terms of Eqs (66-67) is highly non
physical for an exciton-exciton scattering :
As it contains the exciton energy $E_l+E_n$
(or $E_i+E_j$), which is essentially equal to
twice the band gap, such a scattering would
be band gap dependent. We could of course get
rid of this problem by deciding to use for
$H$ an hamiltonian without the band gap i.\
e.\ by dropping $\Delta$ in Eq (3).

Actually the exciton-exciton scattering rate
cannot be related to this matrix element for
more fundamental reasons. In usual problems
dealing with interactions, the hamiltonian
can be written as $H=H_0+V$. The Fermi golden
rule then says that the transition rate
between two different $H_0$ eigenstates
$|i\rangle$ and $|f\rangle$ results from the
possible interaction between these two states
through $|\langle f|V|i\rangle|^2$. When
these states are eigenstates of $H_0$, they
form an orthogonal basis, so that this
transition rate is \emph{also} equal to
$|\langle f|H|i\rangle|^2$, since $\langle
f|H_0|i\rangle=0$ for
$|f\rangle\neq|i\rangle$. All the
difficulties here come from the fact that $H$
cannot be written as $H_0+V$ : There is no
hamiltonian for which the two-exciton states
$|\phi_{ln}\rangle$ and $|\phi_{ij}\rangle$
are the (exact) eigenstates, so that (i)
these two-exciton states are not orthogonal
for $(l,n)\neq(i,j)$ ; (ii) the matrix
elements of $H$ between these states have no
reason to be equal to the matrix element of
``the'' interacting potential $V$ --
which, anyway, cannot be formally
extracted from $H$. Consequently there
is no reason to believe that the
exciton-exciton scattering rate is given
by
$|\langle\phi_{ln}|H|\phi_{ij}\rangle|^2$.

\section{Conclusion}     

We have reconsidered our commutation
technique designed to deal with interacting
close-to-boson particles and extended it to
excitons with spin degrees of freedom.
Although more cumbersome, the expressions of
the two important parameters of this
commutation technique, namely
$\xi_{lnij}^{\mathrm{dir}}$ and
$\lambda_{lnij}$, are still rather
transparent. When the ``spins'' of the
electron and the hole making the exciton are
used -- instead of the total kinetic momentum
of the exciton --, these two parameters
appear as a product of a spin part and an
orbital part which have both a \emph{very
simple physical meaning} (see Eqs (22) and
(25) for $\xi_{lnij}^{\mathrm{dir}}$ and Eqs
(31) and (33) for $\lambda_{lnij}$). The
first one $\xi_{lnij}^{\mathrm{dir}}$
corresponds to all Coulomb interactions
between the $(i,j)$ or $(l,n)$ excitons when,
on both sides, these excitons are made with
the same electron-hole pairs $(e,h)$ and
$(e',h')$. The second one 
$\lambda_{lnij}$ has a purely fermionic
origin and is simply related to the fact that
the $(i,j)$ and $(l,n)$ excitons can be made
with different pairs, $(e,h)$ $(e',h')$ and
$(e,h')$ $(e',h)$. This commutation technique
allows to easily calculate any matrix
elements between exciton states in an
\emph{exact} way. In this paper we have
explicitly calculated the matrix elements of
$H$ between two-exciton states and we have
shown that they cannot be related to the
scattering rate of two excitons as previously
proposed.

\newpage

\hbox to \hsize {\hfill APPENDIX A :
Calculation of the $\gamma_{li}(\mathbf{q})$
\hfill}
\vspace{0.5cm}

Using Eq (17), $\gamma_{li}(\mathbf{q})$ reads
\begin{equation}
\gamma_{li}(\mathbf{q})=\beta_{\nu_l\nu_i}
(\alpha_ha_x\mathbf{q})-\beta_{\nu_l\nu_i}
(-\alpha_ea_x\mathbf{q})\ ,
\end{equation}
with $\beta_{\nu\nu'}(\mathbf{u})$ defined by
\begin{equation}
\beta_{\nu\nu'}(\mathbf{u})=\langle x_{\nu}|
e^{i\mathbf{u}.\mathbf{r}/a_x}|x_{\nu'}
\rangle =\beta_{\nu'\nu}^\ast(-\mathbf{u})\ ,
\end{equation}
and $a_x$ chosen to be the 3D Bohr radius.

From Eq (69), we see that $\beta_{\nu\nu'}
(\mathbf{0})=\delta_{\nu\nu'}$, the value
$\mathbf{u}=\mathbf{0}$ being obtained either
for $\mathbf{q}=\mathbf{0}$ or for
$\alpha_{e,h}=0$, i.\ e.\ $m_h$ (or $m_e$)
infinite.

We can note that, if the $\nu$ and $\nu'$
states have the same parity, $\beta_{\nu\nu'}
(\mathbf{u})=\beta_{\nu\nu'}(-\mathbf{u})$,
so that $\gamma_{li}(\mathbf{q})=0$ for
$m_e=m_h$ : There is no direct scattering
towards a same parity state if the electron
and hole masses are equal.

We can also note that the scattering
$\gamma_{li}(\mathbf{q})$ depends on
$q=|\mathbf{q}|$ only if the $(l,i)$ states
are S states.

Let us now calculate some values of these
$\gamma_{li}(\mathbf{q})$, for 3D and 2D
systems, when the $(l,i)$ states are
S states.

\textbf{1) 3D case}

In 3D, the exciton relative motion wave
function reads
\begin{equation}
\langle \mathbf{r}|x_{\nu}\rangle=a_x^{-3/2}\,
\varphi_{\nu}^{(3D)}(\rho,\theta,\varphi)\ ,
\end{equation}
where $\rho=r/a_x$. The wave functions of the
lowest energy S states are given by
\begin{equation}
\varphi_{1s}^{(3D)}=e^{-\rho}/\sqrt{\pi}\ ,
\hspace{4cm}\varphi_{2s}^{(3D)}=(2-\rho)
e^{-\rho/2}/\sqrt{32\pi}\ . 
\end{equation}

In 3D, Eq (69) reads for S states,
\begin{equation}
\beta_{\nu\nu'}^{(3D)}(u)=\int_0^{+\infty}
\rho^2d\rho
\int_0^{2\pi}d\varphi\int_0^{\pi}\sin\theta\,
d\theta\,e^{iu\rho \cos\theta}\,\varphi_{\nu}
^{(3D)\ast}(\rho,\theta,\varphi)\,\varphi
_{\nu'}^{(3D)}(\rho,\theta,\varphi)\ .
\end{equation}
By inserting Eq (71) into Eq (72), we find
\begin{eqnarray}
\beta_{1s,1s}^{(3D)}(u)&=&\frac{1}{(1+u^2/4)^2}\
,
\\ \beta_{2s,2s}^{(3D)}(u)&=&\frac
{(1-u^2)(1-2u^2)}{(1+u^2)^4}\ ,
\\ \beta_{1s,2s}^{(3D)}(u)&=&\frac
{2^{17/2}\,u^2}{3^6\,(1+4u^2/9)^3}\ .
\end{eqnarray}
From these $\beta_{\nu\nu'}^{(3D)}(u)$, we
can easily deduce the $\gamma_{li}^{(3D)}
(\mathbf{q})$ according to Eq (68), and the
direct scattering $\xi_{lnij}^{\mathrm{dir}}$
according to Eq (23).

\textbf{2) 2D case}

In 2D, the exciton relative motion wave
function reads
\begin{equation}
\langle \mathbf{r}|x_{\nu}\rangle=a_x^{-1}\,
\varphi_{\nu}^{(2D)}(\rho,\varphi)\ .
\end{equation}
The lowest energy S state wave functions are
given by
\begin{equation}
\varphi_{1s}^{(2D)}=\left[2^{3/2}/\sqrt{\pi}
\right]e^{-2\rho}\ ,\hspace{2cm}
\varphi_{2s}^{(2D)}=\left[(2/3)^{3/2}/\sqrt{\pi}
\right](1-4\rho/3)\,e^{-2\rho/3}\ .
\end{equation}

For 2D S states, Eq (69) reads
\begin{equation}
\beta_{\nu\nu'}^{(2D)}(u)=\int_0^{+\infty}\rho
d\rho\int_0^{2\pi}d\varphi\,e^{iu\rho
\cos\varphi}\,\varphi_{\nu}^{(2D)\ast}(\rho,
\varphi)\,\varphi_{\nu'}^{(2D)}(\rho,\varphi)\
.
\end{equation}
Inserting Eq (77) into Eq (78), we find
\begin{eqnarray}
\beta_{1s,1s}^{(2D)}(u)&=&\frac{1}{(1+u^2/16)
^{3/2}}\ ,
\\
\beta_{2s,2s}^{(2D)}(u)&=&\frac{1-(27u^2/16)
+(81u^4/256)}{(1+9u^2/16)^{7/2}}\ ,
\\
\beta_{1s,2s}^{(2D)}(u)&=&\frac{3^{7/2}\,u^2}
{2^9\,(1+9u^2/64)^{5/2}}\ .
\end{eqnarray}
From these $\beta_{\nu\nu'}^{(2D)}(u)$, we
can deduce the
$\gamma_{li}^{(2D)}(\mathbf{q})$ according to
Eq (68) and the direct scattering
$\xi_{lnij}^{\mathrm{dir}}$ according to Eq
(23). For direct scatterings within the 1S
states only, we recover the result given in
ref.\ (11).

\newpage

\hbox to \hsize {\hfill APPENDIX B :
Expressions of
$\hat{\xi}_{lnij}^{\mathrm{dir}}$,
$\hat{\lambda}_{lnij}$ and 
$\hat{\xi}_{lnij}^{\mathrm{right}}$ in
$\mathbf{r}$ space
\hfill}

\vspace{0.5cm}

In order to show the equivalence between Eqs
(24) and (25), it is simpler to start from Eq
(25). Let us consider the $V_{ee'}$ term. By
taking the Fourier transform of $V_{ee'}$,
namely
\begin{equation}
V_{ee'}=\sum_{\mathbf{q}\neq\mathbf{0}}
V_{\mathbf{q}}\,e^{i\mathbf{q}.(\mathbf{r}_e
-\mathbf{r}_{e'})}\ ,
\end{equation}
and the Fourier transform of the relative
motion part of the wave function given in Eq
(26), namely
\begin{equation}
\phi_i(\mathbf{r}_e,\mathbf{r}_h)=\frac{1}
{\mathcal{V}}\sum_{\mathbf{k}_i}
\langle\mathbf{k}_i|x_{\nu_i}\rangle\,
e^{i\mathbf{K}_i^e.\mathbf{r}_e}\,
e^{i\mathbf{K}_i^h.\mathbf{r}_h}\ ,
\end{equation}
with
$\mathbf{K}_i^e=\mathbf{k}_i+\alpha_e
\mathbf{Q}_i$ and $\mathbf{K}_i^h=-
\mathbf{k}_i+\alpha_h\mathbf{Q}_i$, we
can rewrite the
$V_{ee'}$ term of Eq (25) as
\begin{eqnarray}
\hat{\xi}_{ee'}^{\mathrm{dir}}\left(_{n\ j}
^{l\
i}\right)=\sum_{\mathbf{k}_l,\mathbf{k}_n,
\mathbf{k}_i,\mathbf{k}_j}
\langle x_{\nu_l}|\mathbf{k}_l\rangle\,
\langle x_{\nu_n}|\mathbf{k}_n\rangle\,
\langle\mathbf{k}_i|x_{\nu_i}\rangle\,
\langle\mathbf{k}_j|x_{\nu_j}\rangle\,
\sum_{\mathbf{q}\neq\mathbf{0}}V_{\mathbf{q}}
\frac{1}{\mathcal{V}^4}
\int d\mathbf{r}_e\,d\mathbf{r}_h\,
d\mathbf{r}_{e'}\,d\mathbf{r}_{h'}\,
\nonumber
\\ \times
e^{i\left[(-\mathbf{K}_l^e+\mathbf{K}_i^e+
\mathbf{q}).\mathbf{r}_e+(-\mathbf{K}_l^h+
\mathbf{K}_i^h).\mathbf{r}_h+(-\mathbf{K}_n^e
+\mathbf{K}_j^e-\mathbf{q}).\mathbf{r}_{e'}+
(-\mathbf{K}_n^h+\mathbf{K}_j^h).\mathbf{r}_{h'}
\right]}\ .
\end{eqnarray}
The integral being equal to
$\delta_{\mathbf{K}_l^e,\mathbf{K}_i^e+
\mathbf{q}}\,\delta_{\mathbf{K}_l^h,
\mathbf{K}_i^h}\,\delta_{\mathbf{K}_n^e,
\mathbf{K}_j^e-\mathbf{q}}\,
\delta_{\mathbf{K}_n^h,\mathbf{K}_j^h}$, we
immediately recover the first term of Eq (24).

By transforming in the same way the other
terms in $V_{hh'}$, $V_{eh'}$, $V_{e'h}$, it
is easy to derive Eq (24) from Eq (25).

In order to show the
equivalence of Eqs (32) and (33), we can
proceed similarly : By inserting Eq (83) into
Eq (33) and by performing the integral, we
immediately obtain Eq (32).

We do the same to show the equivalence of Eqs
(42) and (46) : We insert Eqs (82,83) into Eq
(46) and perform the integral. This
immediately gives the first term of Eq (42).
The same transformation of the terms in
$V_{hh'}$, $V_{eh'}$, $V_{e'h}$ leads to the
three other terms of Eq (42).

\newpage

\hbox to \hsize {\hfill APPENDIX C : Explicit
calculation of
$F_{1s1s1s1s}(\mathbf{0},\mathbf{0})$

\hfill}

\vspace{0.5cm}

From Eq (36), we get
\begin{equation}
F_{1s1s1s1s}(\mathbf{0},\mathbf{0})=
\sum_{\mathbf{k}}|\langle\mathbf{k}|x_{1s}
\rangle|^4\ .
\end{equation}
The Fourier transforms of the relative motion
wave functions $\langle\mathbf{k}|x_{1s}
\rangle$ are respectively given, in 3D and
2D, by
\begin{eqnarray}
\langle\mathbf{k}|x_{1s}\rangle^{(3D)}&=&\frac
{8\sqrt{\pi}\,a_x^{3/2}}{\sqrt{\mathcal{V}}
(1+a_x^2k^2)^2}\ ,
\\
\langle\mathbf{k}|x_{1s}\rangle^{(2D)}&=&\frac
{\sqrt{2\pi}\,a_x}{\sqrt{\mathcal{S}}\left(
1+\frac{a_x^2k^2}{4}\right)^{3/2}}\ ,
\end{eqnarray}
$a_x$ being the 3D exciton Bohr radius.

In 3D we thus obtain
\begin{equation}
F_{1s1s1s1s}^{(3D)}(\mathbf{0},\mathbf{0})=
\frac{2^{11}\,a_x^3}{\mathcal{V}}\int_0
^{+\infty}\frac{x^2dx}{(1+x^2)^8}
=\frac{33\pi\,a_x^3}{2\,\mathcal{V}}\ .
\end{equation}

In 2D we get
\begin{equation}
F_{1s1s1s1s}^{(2D)}(\mathbf{0},\mathbf{0})=
\frac{2\pi\,a_x^2}{\mathcal{S}}\int_0
^{+\infty}\frac{xdx}{\left(1+\frac{x^2}{4}
\right)^6}=\frac{4\pi\,a_x^2}{5\,\mathcal{S}}
\ .
\end{equation}

\newpage

\hbox to \hsize {\hfill REFERENCES
\hfill}

\vspace{0.5cm}

\noindent
(1) See for instance : H. HAUG and S. KOCH,
Quantum Theory of the Optical and Electronic
Properties of Semiconductors, World
Scientific, London (1990).

\noindent
(2) V.M. AXT, K. VICTOR and A. STAHL, Phys.\
Rev.\ B \underline{53}, 7244 (1996).

\noindent
(3) W. SCH\"{A}FER, D.S. KIM, J. SHAH, T.C.
DAMEN, J.E. CUNNINGHAM, K.W. GOOSSEN, L.N.
PFEIFFER and K. K\"{O}HLER, Phys.\ Rev.\ B
\underline{53}, 16429 (1996).

\noindent
(4) V.M. AXT and S. MUKAMEL, Rev.\ Mod.\
Phys.\ \underline{70}, 145 (1998).

\noindent
(5) T. MEIER and S.W. KOCH, Phys.\ Rev.\ B
\underline{59}, 13202 (1999).

\noindent
(6) See for instance : A. KLEIN and E.R.
MARSHALEK, Rev.\ Mod.\ Phys.\
\underline{63}, 375 (1991).

\noindent
(7) See for instance : H. HAUG and S.
SCHMITT-RINK, Progr.\ Quant.\ 
Electron.\ \underline{9}, 3 (1984) ; more
precisely their Eqs (5.\ 7,9).

\noindent
(8) M. COMBESCOT and O. BETBEDER-MATIBET,
Europhys.\ Lett.\ \underline{58}, 87 (2002).

\noindent
(9) Recently, a ``kinetic correction'' has
been added to the usual boson-exciton
effective hamiltonian by G. ROCHAT, C. CIUTI,
V. SAVONA, C. PIERMAROCCI and A. QUATTROPANI
(Phys.\ Rev.\ B \underline{61}, 13856 (2000)).
It may look like a fermionic term. It is
however intrinsically incorrect as it depends
on free electron and free hole energies
($E_k^c$ and $E_k^v$ with their notations),
while it should depend on exciton energies
$E_i$ (see our Eq (56)).

\noindent
(10) A recent work by S. BEN-TABOU DE LEON and
B. LAIKHTMAN (Phys.\ Rev.\ B \underline{63},
125306-1 (2001)) uses a new bosonisation
procedure. It is still incorrect as it relies
on a separation between a subspace with all
the excitons in the ground state and a
subspace with at least one exciton in an
excited state. Such a separation is
meaningless due to our Eq (34).

\noindent
(11) C. CIUTI, V. SAVONA, C. PIERMAROCCI, A.
QUATTROPANI and P. SCHWENDIMANN, Phys.\ Rev.\
B \underline{58}, 7926 (1998).

\noindent
(12) J. M. LUTTINGER and W. KOHN, Phys.\
Rev.\ \underline{97}, 969 (1955).

\noindent
(13) J. M. LUTTINGER, Phys.\ Rev.\
\underline{102}, 1030 (1956).

\noindent
(14) M. COMBESCOT and P. NOZIERES, Solid
State Com.\ \underline{10}, 301 (1972).

\noindent
(15) In 2D, and in the particular case
$\nu_l=\nu_n=\nu_i=\nu_j=1S$, by setting
$\mathbf{Q}_l-\mathbf{Q}_i=\mathbf{q}$, we
can check that $\hat{\xi}^{\mathrm{dir}}$ is
nothing but the quantity $H_{dir}(q)$
introduced by Ciuti et al $^{(11)}$ in their
Eq (12). By inserting our Eqs (68) and (79)
into our Eq (23), we recover the result they
give in their Eqs (19-20).

\newpage

\hbox to \hsize {\hfill FIGURE CAPTIONS
\hfill}

\vspace{0.5cm}

\noindent
Fig.\ (1)

Direct Coulomb scattering of an $i$ exciton
into an $l$ exciton, while a $j$ exciton is
scattered into an $n$ exciton. As Coulomb
interaction conserves the angular momenta of
electrons and holes, we must have $s_l=s_i$,
$s_n=s_j$, $m_l=m_i$, $m_n=m_j$. As it also
conserves momenta, we must have
$\mathbf{Q}_i+\mathbf{Q}_j=\mathbf{Q}_l+
\mathbf{Q}_n$, between the centers of mass
of the ``in'' and ``out'' excitons. Finally
this direct Coulomb scattering contains the
quantities $\gamma_{li}(\mathbf{q})$ and
$\gamma_{nj}(-\mathbf{q})$, with $\mathbf{q}=
\mathbf{Q}_l-\mathbf{Q}_i$, which
characterize the scattering of a $\nu_i$
exciton into a $\nu_l$ state and a $\nu_j$
exciton into a $\nu_n$ state, under a
$\mathbf{q}$ excitation (see Eq (17)).

\noindent
Fig.\ (2)

Bare exchange coefficient : The $i$ and $l$
excitons are made with the same electron but
different holes ; we do have $s_l=s_i$ but
$m_l=m_j$.

\noindent
Fig.\ (3)

The dimensionless exchange coefficient
$\Lambda(p)$, given by Eq (38), as a
function of $P=p\lambda_{2D}$, for 2D
excitons and three different values of
$m_e/m_h$. Thick solid line : $m_e/m_h=0$.
Thin solid line : $m_e/m_h=0.5$. Dashed line
: $m_e/m_h=1$.

\noindent
Fig.\ (4)

Right exchange Coulomb scattering (a), and
left exchange Coulomb scattering (b). In the
right exchange scattering, the exchange of
holes making the $(i,j)$ excitons is made
\emph{before} the direct Coulomb interaction
which then scatters the $(p,r)$ excitons into
$(l,n)$ states. Note that the electron-hole
interactions are \emph{between} the $(l,n)$
excitons but \emph{inside} the $(i,j)$
excitons. Exchange and Coulomb processes
conserving spins, we do have $s_l=s_i$,
$s_n=s_j$ along with $m_l=m_j$, $m_n=m_i$,
the holes being crossed in the process. Note
that the $(l\leftrightarrow n)$ change
appearing in the definition of
$\xi_{lnij}^{\mathrm{dir}}$ and
$\lambda_{lnij}$ generates a similar term in
which the electrons are crossed instead of
the holes. This restores the electron-hole
symmetry.

\noindent
Fig.\ (5)

The dimensionless difference between left and
right exchange scatterings $\Delta g(p)$,
given by Eq (57), as a function of
$P=p\lambda_{2D}$, for 2D excitons and three
different values of $m_e/m_h$. Thick solid
line : $m_e/m_h=0$. Thin solid line :
$m_e/m_h=0.5$. Dashed line : $m_e/m_h=1$.

\end{document}